\documentclass[useAMS,usenatbib]{mn2e}
\usepackage{epsfig,amsmath,amssymb,amsfonts,mathrsfs,latexsym,graphicx}

\usepackage{fixltx2e}
\usepackage{amsmath}
\usepackage{float}
\usepackage[caption = false]{subfig}
\usepackage{multirow}
\include{journals}

\graphicspath{{images/}}

\newcommand{\od}{$\Omega_{\mathrm{dust}}$}

\def\lsim{\raise0.3ex\hbox{$<$}\kern-0.75em{\lower0.65ex\hbox{$\sim$}}}

\newcommand{\apj}{ApJ}

\newcommand{\aj}{AJ}
\newcommand{\aap}{A\&A}
\newcommand{\mnras}{MNRAS}

\newcommand{\pasp}{PASP}

\newcommand{\jcap}{JCAP}

\newcommand{\araa}{ARA\&A}

\title[Cosmic transparency]{\boldmath The cosmic transparency measured with Type Ia supernovae: implications for intergalactic dust}
\author[Goobar, Dhawan \&  Scolnic]{Ariel Goobar$^1$, Suhail Dhawan$^1$, Daniel Scolnic$^2$
\\ 
$^1$ The Oskar Klein Centre, Physics Department, Stockholm
  University, SE 106 91 Stockholm, Sweden \\
$^2$ Kavli Institute for Cosmological Physics, University of Chicago, Chicago, IL 60637, USA}
\begin{document}

\date{Accepted ... Received ...; in original form ...}

\pubyear{2018}

\maketitle
\begin{abstract}
Observations of high-redshift Type Ia supernovae (SNe~Ia) are used to study the cosmic transparency at optical wavelengths. Assuming a flat $\Lambda$CDM cosmological model based on BAO and CMB results, redshift dependent deviations of SN~Ia distances are used to constrain mechanisms that would dim light. The analysis is based on the most recent Pantheon SN compilation, for which there is a $0.03\pm0.01 {\textrm \,(\rm stat)}$ mag discrepancy in the distant supernova distance moduli relative to the $\Lambda$CDM model anchored by supernovae at $z<0.05$.  While there are known systematic uncertainties that combined could explain the observed offset, here we  entertain the possibility that the discrepancy may instead be explained by scattering of supernova light in the intergalactic medium (IGM). We focus on two effects: Compton scattering by free electrons and extinction by dust in the IGM. We find that if the discrepancy is due entirely to dimming by dust, the measurements can be modeled with a cosmic dust density $\Omega_{\rm IGM}^{\rm dust} = 8 \cdot 10^{-5} (1+z)^{-1}$,  corresponding to an average attenuation of $2\cdot 10^{-5}$ mag Mpc$^{-1}$ in V-band. Forthcoming SN~Ia studies may provide a definitive measurement of the IGM dust properties, while still providing an unbiased estimate of cosmological parameters by introducing additional parameters in the global fits to the observations.
\end{abstract}

\section{Introduction}\label{sec-intro}
Observations of Type Ia supernovae (SNe~Ia) led to the discovery of the accelerated expansion of the universe \citep{1998AJ....116.1009R,Perlmutter:1998np}, see \cite{2011ARNPS..61..251G} for a review of the field. Current compilations of SNe~Ia data-sets \citep{Betoule2014,2017arXiv171000845S} are critical to constrain the properties of the recent period of accelerated expansion and distinguish between different explanations of late-time acceleration. No significant deviation has been seen so far from the standard cosmological model, flat $\Lambda$CDM, see e.g.,  \citep{2017JCAP...07..040D,2017arXiv170700603L,2017arXiv170507336L}. 

Early attempts to explain the observed dimming of the high-redshift SNe~Ia by unresolved astrophysical effects invoked (among other things) extinction by large dust grains (`grey dust') in the intergalactic medium (IGM) that would have escaped the reddening corrections applied to the data \citep{1999ApJ...525..583A,2001ApJ...560..599A}. Techniques to  explore this possibility were proposed by \citep{2002A&A...384....1G}, and analysis that followed showed that dimming by IGM dust was inadequate as an alternative explanation for the apparent accelerated expansion of the universe. For example, 
the study of discrete sources detected by SCUBA account for nearly all the emission at 850$\mu$m, leaving little room in the observed far-infrared background for diffuse emission from sufficiently large large amounts of dust in the IGM \citep{2001ARA&A..39..249H}. Constraints on cosmic opacity were obtained from high-redshift quasar colours \citep{2003JCAP...09..009M}, and including soft X-ray sources \citep{2009MNRAS.397.1976D,2012MNRAS.426.3360J},  and also from tests of the relation between the luminosity and angular diameter distances \citep[distance duality relation;][]{1933PMag...15..761E} as tested in \citet{2009ApJ...696.1727M,2009JCAP...06..012A,2012JCAP...12..028N}. %
When analyzing the impact on previous SNe~Ia data-sets, \citet{2006MNRAS.372..191C} and \citet{2010MNRAS.406.1815M} concluded that the potential systematic effect from  dimming by intergalactic dust, $\delta m$, on the inferred equation of state parameter of the dark energy equation-of-state $w$ was a few percent, $\delta w \sim 2\delta m$, and thus (still) sub-dominant, but a potential concern for future precision cosmology surveys based on SNe~Ia. 

In this work, we use the flux independent estimate of $\Omega_M$ in the standard flat $\Lambda$CDM cosmological model from the combination of baryon acoustic oscillations (BAO) and cosmic microwave background measurements (CMB). The derived redshift dependent
luminosity distance predictions are compared with measurements using SNe~Ia  in \cite{2017arXiv171000845S}. Differences between the two are used to study 
the cosmic transparency and to explore the possibility that light is attenuated in the IGM. Dimming of light along the line of sight would lead to an excess in the measured distance modulus from SNe~Ia, but not influence the BAO and CMB results. In particular, we focus on two effects that could lead to light losses affecting the SN distances: Compton scattering by free electrons and extinction by dust in the IGM. While the cosmic density of electrons is well characterized by the baryon density measurements from the CMB, the presence of dust in the IGM remains poorly constrained.

From the integrated star formation rate,  \citet{2011arXiv1103.4191F} estimates the total amount of dust produced by stellar evolution  over cosmic time to be \od\, $\sim 10{^{-5}}$, in units of present day critical mass density. This estimate of the total dust density  appeared to be a factor of a few larger than the amount of dust observed in galactic disks \cite{2004ApJ...616..643F,2007MNRAS.379.1022D}, motivating the exploration of other locations for dust grains, e.g. in the IGM. Theoretical studies indicate that dust grains can be efficiently transported from disks to the IGM through winds and radiation pressure \citep{1999ApJ...525..583A,2001ApJ...560..599A,2005MNRAS.358..379B}. Observational evidence of dust extending beyond galaxies has been established using low-redshift foreground/background galaxy superposition to detect extinction up to five times the optical extent of the spiral galaxies \citep{2009AJ....137.3000H}. \emph{Herschel} observations of the nearby galaxy M82 shows emission from cold dust up to 20 kpc from the center of the galaxy  \citep{2010A&A...518L..66R}, and an excess reddening signal on scales ranging from 20 kpc to a few Mpc was found in cross-correlation studies between the colours of distant quasars and foreground galaxies as a function of the angular separation of the galaxies \cite{2010MNRAS.405.1025M}.
More recently, constraints on the amount of intergalactic dust have been inferred from Mg II absorbers \citep{2012ApJ...754..116M} to be $\Omega_{\mathrm{\rm IGM}}^{\rm dust} \sim 2 \cdot 10^{-6}$, which is about half of the total dust content inferred to exist outside galaxies \citep{2011arXiv1103.4191F}.

Here, we aim at measuring the cosmic transparency directly and place constraints on the density of dust in the IGM using the SN~Ia magnitude-redshift relation, where the cosmological model has been determined from independent probes not susceptible to extinction. Besides their relevance for cosmological distances estimates, these estimates are of interest to evaluate the impact of dust on CMB polarization experiments \citep[e.g., see][]{2018ApJ...853..127H}, and for precision measurements of CMB spectral distortions \citep{2016ApJ...825..130I}. We also compare the measured transparency with the expectations of light attenuation due to Compton scattering on free electrons in the intergalactic medium, see e.g. \cite{2008ApJ...682..721Z,2017ApJ...836..107H}. 

The methodology and data are presented in Section~\ref{sec-data}, we describe light attenuation by intergalactic dust in Section~\ref{sec-light} and opacity from Compton scattering in Section~\ref{sec-opacity}. We present our results in Section~\ref{sec-results} and a discussion in Section~\ref{sec-disc}.  

\section{Methodology and data}\label{sec-data}
For our analyses we use the latest compilation of SNe~Ia from the \texttt{Pantheon} sample \citep{2017arXiv171000845S}, including 1049 SNe~Ia extending out to a maximum redshift of 2.26.

We calculate the residuals in the Hubble diagram, $\Delta\mu$, relative to a fiducial cosmology. The expression for the residuals is given by 
\begin{equation}
\Delta \mu =  m_B + \alpha \cdot x_1 - \beta \cdot c + \Delta_M + \Delta_B - M_B - \mu 
\label{eq:hub_res}
\end{equation}

where $m_{B}$ is the  observed apparent peak magnitude. This is corrected for the light curve width ($x_1$), colour ($c$) with coefficients $\alpha$ and $\beta$ and host galaxy properties ($\Delta_M$).  \citet{2017ApJ...836...56K} account for biases due to intrinsic scatter and selection effects and use the BEAMS with bias correction method (BBC). BBC simultaneously fits for $\alpha$ and $\beta$ from equation~\ref{eq:hub_res}. The method relies on \citet{2011ApJ...740...72M} along with extensive simulations to correct SALT2 parameters, $m_B$, $x_1$ and $c$. This predicted bias from  simulations is expressed as $\Delta_B$  \citep[see][for details]{2017arXiv171000845S}.   We note that the colour correction $\beta \cdot c$ accounts for a combination of the intrinsic colour of the SN and the extinction from the host galaxy dust \citep[see, e.g.][]{2007A&A...466...11G} but does not explicitly account for an extinction from intervening matter between the SN redshift and the observer. Milky-Way extinction is accounted for separately by \cite{2017arXiv171000845S} using the extinction map from \cite{2011ApJ...737..103S}.  
$M_B$ is the absolute peak magnitude and  $\mu$ is the distance 
modulus given by:
\begin{equation}
\mu = 5 \mathrm{log}_{10}(D_L) + 25
\label{eq:dist_mod}
\end{equation}

and the luminosity distance $D_L$, which for a fiducial flat $\Lambda$CDM cosmology is given by

\begin{equation}
D_L = \frac{c (1+z_s)}{H_0} \int_0^{z_s} \frac{d\mathrm{z}}{\sqrt{\Omega_\mathrm{M} (1+z)^{3} + 1 - \Omega_\mathrm{M}}}.
\label{eq:lum_dist}
\end{equation}

Since the presence of intergalactic dust would only affect the inferred luminosity distance from SNe~Ia and not the angular scale of the Baryon Acoustic Oscillations \citep[BAO;][]{2017MNRAS.470.2617A} or the cosmic microwave background \citep[CMB;][]{Planck2015}, we use the BAO and CMB data to evaluate the best fit cosmological parameters. Assuming a flat $\Lambda$CDM model, the present day matter density ($\Omega_M$) determines the redshift dependence of the SN~Ia Hubble diagram, whereas $M_B$ and $H_0$ determine the absolute scale. Using the compressed CMB likelihood from the \emph{Planck} satellite \citep{Planck2015} and BAO observation at $z_{\mathrm{eff}} = 0.106, 0.15, 0.32, 0.57$ from \citet{2011MNRAS.416.3017B, 2015MNRAS.449..835R, 2014MNRAS.441...24A}, we get a best fit $\Omega_M = 0.309 \pm 0.006$ and hence this is the adopted cosmological model in this work.

\begin{table}
\begin{center}
\caption{The Hubble residuals for the \texttt{Pantheon} sample relative to a flat $\Lambda$CDM cosmology with $\Omega_M =0.309$ and $\mathcal{M} = -28.59$. }
\begin{tabular}{|c|c|c|c|c|}
\hline\hline
z$_{bin}$ & $\Delta \mu$ & {$\sigma_{\Delta \mu}$} & $N_{SN}$ & Light-path \\
          & [mag]  & [mag]  & & [Gpc] \\  
\hline
0.125 & 0.035 & 0.008 & 254 & 0.5 \\
0.3 & 0.029 & 0.007 & 355 & 1.1 \\
0.5 & 0.044 & 0.013 & 118 & 1.5 \\
0.7 & 0.037 & 0.015 & 80 & 1.9 \\
0.9 & 0.044 & 0.019 & 59 & 2.2 \\
1.1 & 0.018 & 0.069 & 7 &  2.5 \\
1.3 & 0.044 & 0.049 & 10 & 2.7 \\
1.83 & -0.03 & 0.051 & 6 & 3.0 \\
\hline 
\end{tabular}
\label{tab:residual}
\end{center}
\end{table}

Since $M_B$ and $H_0$ are degenerate, we evaluate a combination of the two parameters $\mathcal{M} = M_B - 5\, \mathrm{log}_{10}(H_0)$. We only use the subset of supernovae with $z<0.05$ to fit $\mathcal{M}$, 
since that is a redshift range where the expected opacity in the intergalactic medium is small enough to be 
neglected, as described in the next section. Based on 160 supernovae, we find $\mathcal{M} = -28.59 \pm 0.01$, at a median redshift $z=0.0265$. In particular, we choose the upper range of the first bin, $z<0.05$, since it includes enough objects to yield a 0.01 mag uncertainty on ${\cal M}$, while the expected attenuation from all the models considered was below that value. Extending the bin to higher redshifts would systematically bias the estimate of the attenuation above the statistical uncertainty. 

We bin the residuals with a $\Delta z$ of 0.2 up to a redshift of 1.4. The first bin excludes the set of supernovae ($z<0.05$) used to fit $\mathcal{M}$. Since the number of SNe at $z > 1.4$ is small, we have only one redshift bin centered at $z = 1.83$. The residuals for each redshift bin, along with the errors, the central redshift, the number of SNe in each bin  and the corresponding light-travel path are shown in Table~\ref{tab:residual}, where it can be seen that there is a statistically significant offset in the Hubble residuals, $\Delta \mu$. This suggests the existence of potential systematic effects in the low redshift data, see \cite{2017arXiv171000845S} for a discussion. Alternative explanations may involve a redshift dependence of the SNIa brightness, or dimming of light in the line of sight, since the offset is such that the more distant supernovae are fainter than the $z<0.05$ sample. In the next section, we examine the latter. In particular, we compare the observed offset in the Hubble residuals to the expected attenuation by dust in the intergalactic medium.

\section{Light attenuation by intergalactic dust}
\label{sec-light}
The dimming of supernova light along the line of sight by scattering on dust grains in the IGM depends on the grain properties, the redshift distribution of the dust density and the physical path length. The latter involves the expansion rate of the universe, $H(z)$.
In particular, the  wavelength dependent optical depth, $\tau_\lambda$,  for a source at redshift $z_s$ due to attenuation by a homogeneous density of dust in the IGM, $\rho_{\rm IGM}^{\rm dust}(z)$ is expressed as:
\begin{equation}
\tau_\lambda = 
c \int_{0}^{z_s}
{
{\kappa \left(\lambda \left(\frac{1+z_s}{1+z}\right), R_V\right)
\rho_{\rm IGM}^{\rm dust}(z)} \over {H(z)(1+z)
}
}dz,
\end{equation}
where $\kappa$ is the wavelength dependent mass absorption coefficient (or opacity), reflecting the nature of the dust size and composition. 
As a reference value, and to conform with the study of \cite{2012ApJ...754..116M}, we adopt the SMC  value $\kappa_V \approx 1.54\cdot 10^4$ cm$^2$g$^{-1}$ from
\citep{2001ApJ...548..296W}, a factor of $1.8$ larger than for Milky-Way type dust. 
The wavelength dependence is otherwise parametrized through the total-to-selective coefficient $R_V$, as used for Milky-Way dust \citep{1989ApJ...345..245C,1999PASP..111...63F}.
Normalizing to the present day critical density ($\rho^{c}_0$) and assuming that the IGM dust density scales with redshift as
$\rho^{\rm dust}_{\rm IGM}(0) \cdot (1+z)^\gamma$, we arrive at the expected attenuation:
\begin{align}
A_\lambda^{rest} & = 1.086 \tau_\lambda  \nonumber \\ 
& = 1.086 \cdot \frac{3 c H_0}{8 \pi G}\Omega_{\rm IGM}^{\rm dust} {\cal I}(z_s, \lambda, R_V) \label{eq:grey1} \\
{\rm where} & \nonumber \\ 
{\cal I} &= \int_{0}^{z_s}
{
 {\kappa \left(\lambda \left(\frac{1+z_s}{1+z}\right), R_V\right)
 (1+z)^{\gamma-1} \over \sqrt{\Omega_\mathrm{M} (1+z)^{3} + 1 - \Omega_\mathrm{M}}
}dz}.
\label{eq:grey2}
\end{align}
As the observations are not monochromatic, the comparison should include the convolution of the spectral energy distribution of SNe Ia over the restframe range used for the cosmological parameter estimations, centered around restframe $B$-band. E.g., the highest redshift sources are observed through near-IR filters centered at $0.44\cdot (1+ z_s)$ $\mu$m.  To do this calculation we use the 
\texttt{SNOC} simulation package \citep{2002A&A...392..757G}. As in \cite{2002A&A...384....1G}, we have considered a wide variety of dust models with $R_V$ ranging from 3 to 9. The grey dust limit $(R_V \gg 3)$ is particularly interesting, as it otherwise evades detection based on colours.

\section{Opacity from Compton scattering}
\label{sec-opacity}
The impact of light attenuation by Compton scattering of SN photons on free electrons has been discussed in \cite{2008ApJ...682..721Z,2017ApJ...836..107H} and references therein. As the universe is re-ionized at $z = 6 - 10$, a cosmic density of free electrons that can scatter SN light is expected. The electron density is tied to the baryon density, as it originates primarily from ionization of hydrogen and helium atoms. f Disregarding the fraction of baryons locked to galaxies, we can estimate an upper bound on the electron density in the IGM at ($z=0$), $n_{e} = n_{H} + 2\cdot n_{He}$, where:
\begin{align*}
n_H  & = (1-Y) \cdot \Omega_B \rho^{c}_0/m_p  \\
n_{He} & = \, Y  \cdot \Omega_B \rho^{c}_0/4m_p, \\
\end{align*}
where
$Y$  is the  helium abundance (by mass), $\Omega_B$ and $m_p$ are the comoving baryon density and
the proton mass respectively. 
For optical wavelengths the interaction is well-described by the Thomson cross-section, $\sigma_T \approx 6.625\cdot 10^{-25}$ cm$^{2}$. For redshifts well below the re-ionization limit, the free-electron density is expected to scale inversely proportional to volume, i.e., as $(1+z)^3$, which leads to the attenuation:
\begin{equation}
A = 1.086 \cdot \frac{c \sigma_{T} n_{e}}{H_0} \int_{0}^{z_s} \frac{(1+z)^2}{\sqrt{\Omega_M(1+z)^3 +1 - \Omega_M}} dz,
\label{eqn:compton}
\end{equation}
independent of wavelength, in the regime considered here.
In the Eqn (\ref{eqn:compton}) we adopt standard cosmology, as in the rest of this work and have adopted $\Omega_B h^2 = 0.02225$ and $Y=0.2467$ \citep{Planck2015}, leading to an estimate $n_e = 3.14 \cdot 10^{-7}$ cm$^{-3}$.

\section{Results}\label{sec-results}
\begin{figure*}
\centering
\includegraphics[width=0.9\hsize]{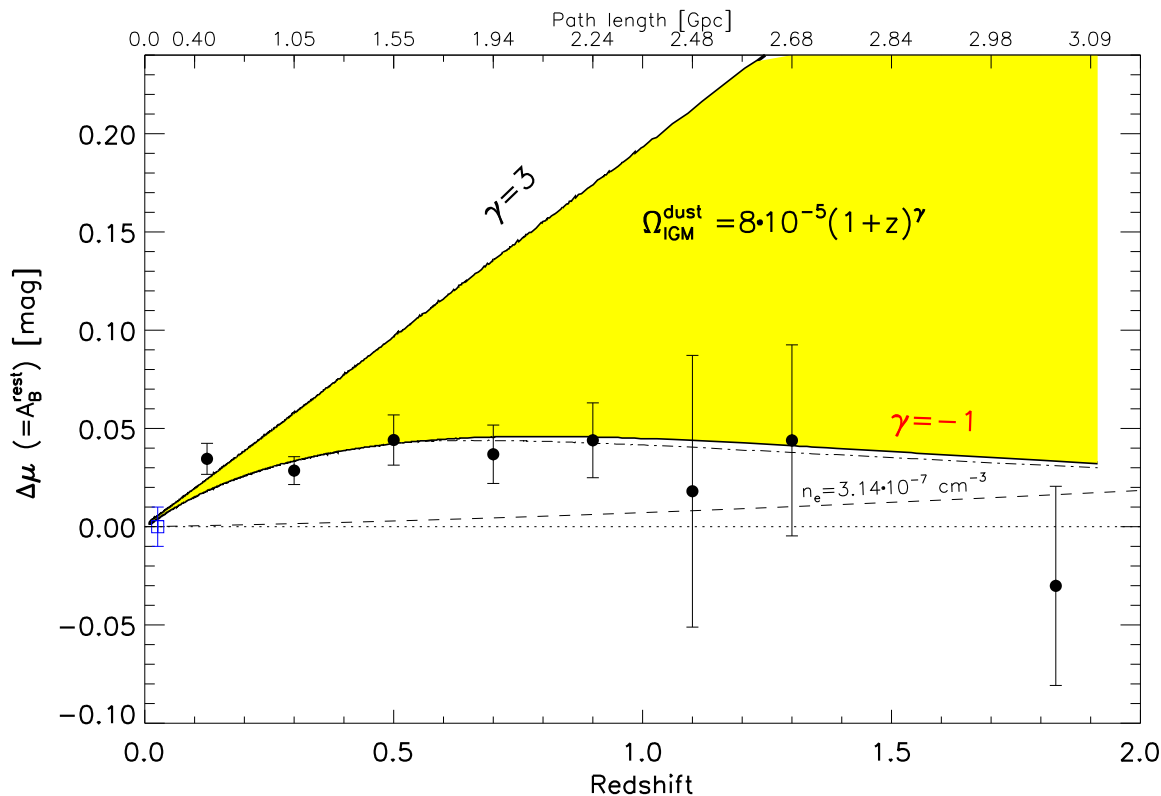}
\caption{Observed SN Ia Hubble residuals, $\Delta \mu$, from the best fit cosmology derived from an independent combination of BAO and CMB data, see also Table 1. The open square symbol indicates the redshift bin used to estimate ${\cal M}$, i.e., the zero level of the Hubble residuals. The shaded region shows the expected attenuation from intergalactic dust parameterized by Fitzpatrick (1999). We consider a wide range of redshift evolutions, all corresponding to a density $\Omega_{\rm IGM}^{\rm dust} = 8 \cdot 10^{-5} (1+z)^\gamma$.  The best fit is found for $\gamma=-1$, indicated by the lower solid line. The steep redshift dependence predicted for  larger (positive) values of $\gamma$ is inconsistent with the observations.  The IGM dust model shown with solid lines corresponds to $R_{V}=9$, while the dot-dashed curve corresponds to $R_{V} =3$, i.e., the model predictions are only mildly affected by the choice on the total-to-selective extinction, as explained in the text. The dashed line shows the predicted attenuation from Compton scattering by free electrons in the IGM, Eqn (\ref{eqn:compton}), well below the accuracy of the Hubble residuals in the \texttt{Pantheon} data \citep{2017arXiv171000845S}. The upper axis scale indicates the light travel-path. The modeled attenuation decreases slightly towards higher redshifts for the best fit ($\gamma=-1$), in spite of the longer path, since these observations are carried out through near-IR filters, where the cross-section for light scattering is smaller than for optical wavelengths.}
\label{fig:HR}
\end{figure*}

If we attribute the difference in the Hubble residuals in Table 1 to be exclusively from attenuation by intergalactic dust, we find that it fits well with $\Omega_{\rm IGM}^{\rm dust}(z) = 8 \cdot 10^{-5} (1+z)^{-1}$, as shown in Figure \ref{fig:HR}. The best fit is found for $\gamma=-1$,
which agrees with the empirically derived density evolution for MgII absorbers. The figure also shows that the model prediction around the best fit value are only very mildly dependent on the assumption on $R_V$, since we have adopted a fixed value for  $\kappa_V$ in all cases,  corresponding to an attenuation of  $2 \cdot 10^{-5}$ mag Mpc$^{-1}$ in V-band, averaged over the redshift range considered. Furthermore, the modeled attenuation is slightly smaller for the highest
redshifts considered,  in spite of the longer light-travel path, since these observations are carried out at near-IR wavelengths, where the expected cumulative attenuation is smaller than in the optical.

For larger values of  $\gamma$, a steeper redshift evolution is expected than what is observed. Figure \ref{fig:HR} also shows the expected attenuation from Compton scattering on free-electrons in the IGM, more than an order of magnitude smaller than the observed offsets.
In this analysis we have neglected the potential scenario where the colour corrections applied to the data could already correct for some small fraction of the reddening in the line of sight, especially if the dust is located at redshifts not very different from the SN host galaxy. While potentially a source of systematic underestimation of $\Omega^{\rm dust}_{\rm IGM}$, the approximation is justified for the present analysis since the colour excess predicted by the models considered are small, $E(B-V) \lsim 0.01$ mag, in comparison to the intrinsic colour variations of SNe~Ia, $\sigma_{(B-V)} \sim 0.05$ mag \citep{2008A&A...487...19N}. Moreover, the correlation with the applied $\beta\cdot c$ correction is weakened, since most of the expected dimming in the preferred model happens at lower $z$, irrespective of the redshift of the SN. We note that future analysis aiming at extracting an unbiased estimate of $\Omega^{\rm dust}_{\rm IGM}$  could avoid this potential source of error  by fitting the IGM dust model simultaneously with the cosmological parameters and nuisance parameters.  

\section{Discussion}\label{sec-disc}
We have performed the first direct measurement of the cosmic transparency of the universe at restframe optical ($B$-band) using the latest compilation of Type Ia supernovae used in cosmology \citep{2017arXiv171000845S}. We find systematic 
offsets in the Hubble residuals that may be due to calibration and selection effects in the low-z data \citep{Scolnic14a,Scolnic14b,Scolnic15,2017arXiv171000845S}. The residuals exceed the
expectations for attenuation of supernova light by scattering on free electrons in the IGM. However, a model with dust in the IGM following  $\Omega_{\rm IGM}^{\rm dust}(z) =  8 \cdot 10^{-5} (1+z)^{-1}\left(\frac{1.54\cdot 10^4}{\kappa_v [{\rm cm}^2 {\rm g}^{-1}]}\right)$ provides a good statistical fit to the observations.
The found density in this analysis exceeds what has been inferred by \cite{2012ApJ...754..116M} by about an order of magnitude.  There are, however, some important differences between the two measurements. \cite{2012ApJ...754..116M} measure reddening of quasars shining through
MgII absorbers, which are predominantly found close to galaxy halos. Furthermore, since their method is based on measuring colour excess, it is insensitive to grey dust caused by large dust grains. However, theoretical modelling of dust in the IGM by \citet{2018arXiv180204027A} suggests that the density of large grains in the IGM should be close to an order of magnitude larger than small grains. A dominant contribution of large grains would result in a much weaker wavelength dependence in the attenuation than what was found by \cite{2012ApJ...754..116M}. It is therefore possible that the difference between the two measurements could be reconciled once the spatial and grain size sensitivity is accounted for, since the measurement presented in this work is insensitive to the spatial distribution of the IGM dust, and only moderately sensitive to grain size, and in the opposite direction as the QSO result. However, the measured dust density in our analysis remains high compared to theoretical expectations in \citep{2018arXiv180204027A}.
We use the best fit values for the intergalactic dust and find that, if not corrected for, the inferred value of $w$ changes by a $\Delta w \sim 0.05$, i.e., comparable to the quoted statistical uncertainty in \cite{2017arXiv171000845S}. However, if dust in the IGM is confirmed by new data-sets to be present at levels comparable to what has been found here we advocate a different strategy. A correction that renders the cosmological analysis of future larger data-sets unbiased can be achieved by including two more parameters in the global fit, corresponding to $\Omega_{\rm IGM}^{\rm dust}$ and $\gamma$ in Eqns. (\ref{eq:grey1}) and (\ref{eq:grey2}). For the present data-set we found $w = -0.993 \pm 0.073$, i.e., resulting in a somewhat larger uncertainty on the dark energy equation of state parameter in a joint fit between SNe~Ia, BAO and CMB, fixing $\gamma=-1$.

As mentioned, the observations could also be explained by selection effects and calibration systematics between the low-$z$ and high-$z$ SNe~Ia sample. Since the low-$z$ SN anchor is constructed from a variety of SN programs, which combines different photometric systems as well as volume-limited with magnitude-limited surveys, having a homogeneous low-$z$ sample would be important to test the presence of systematics.   Many of these issues are being addressed by the Foundation Survey \citep{2018MNRAS.475..193F}, a follow-up survey of low-z SNe discovered by `rolling surveys' like ATLAS \citep{ATLAS2011} and ASAS-SN \citep{ASASSN2014}, which should observe up to 800 SNe.  A definitive answer may be provided by the Zwicky Transient Facility (ZTF) \citep{2017NatAs...1E..71B}, a new multi-colour rolling survey that between 2018 and 2020 will find $\sim$ 2000 SNe in the Hubble flow out to a redshift of 0.08, an order of magnitude greater than the number of SNe in the same redshift range in the current compilation, implying an improvement by $\sim$ a factor of 3 on the statistical uncertainty, with a 1\% target for systematic errors. 

These low-z samples will be complemented by the Wide-Fast-Deep survey of LSST \citep{LSST2012}, which will provide about an order of magnitude larger statistics on the number of high-z SNe in the same redshift range in the current compilation.  Exploration of even higher redshifts, potentially interesting for modelling of IGM dust enrichment, may become feasible with the next generation space telescope, JWST. The new sets of SNe Ia data will allow us to explore cosmic transparency very accurately, an important aspect for precision cosmology.   



\section*{Acknowledgements}
The authors are grateful 
to Edvard M\"ortsell and Brice M\'enard for stimulating discussions. Special thanks to Garrelt Mellema for clarifying discussions on the estimates of $n_e$. Funding from the Swedish Research Council, the Swedish Space Board and the K\&A Wallenberg foundation made this research possible.   D.S. supported in part by the Kavli Institute
for Cosmological Physics at the University of Chicago
through grant NSF PHY-1125897 and an endowment
from the Kavli Foundation and its founder Fred Kavli.
D.S. is also supported by NASA through
Hubble Fellowship grant HST-HF2-51383.001 awarded
by the Space Telescope Science Institute, which is operated
by the Association of Universities for Research
in Astronomy, Inc., for NASA, under contract NAS 5-
26555.


\end{document}